\documentstyle[preprint,aps,psfig,floats]{revtex}

\begin{document}

\draft

\title{Semiclassical stability 
analysis of a two-photon laser including spatial variation of the cavity field
}

\author{ M. Abdel-Aty$^{1,}$\footnote{Corresponding author:
 abdel-aty@usa.net, abdelaty@hotmail.com, Fax. No. ++2 +93601950} 
 and A.-S F. Obada$^{2}$}

\address{$^{1}$                         
Mathematics Department, Faculty of Science, South Valley University, 
 82524 Sohag, Egypt\\
$^{2}$ Mathematics Department, Faculty of Science, Al-Azhar University, 
 Cairo, Egypt
}

\date{\today}

\maketitle

\begin{abstract}
We investigate the dynamics of a two-photon laser under 
conditions where the spatial variation 
of the cavity field along the cavity axis is important.
The model assumes pumping to the upper state of the two-photon 
transition. We consider the Maxwell-Bloch equations on the 
basis of which we study the stability analysis of the steady state of the system.
The system  is taken to be contained in a  ring-laser cavity.
Asymptotic expansion of the eigenvalue  and analytic information are obtained in some 
realistic limits, such as very large reflectivity, very small cavity losses, or very small population relaxation rate.
The results are illustrated with an application to a specific atomic system 
(Potassium) as an amplifying medium.
\end{abstract}

{\bf PACS:} 
{42.55.Ah}\quad  {General laser theory},   
{42.65.Sf}\quad {Dynamics of nonlinear optical systems; optical instabilities;
optical spatio-temporal dynamics},
{42.25.Ja}\quad{Polarization}


\newpage

\section{Introduction}
        
One of the main obstacles in the realization of the standard form 
of the two-photon laser is a technological one; 
namely the difficulty in constructing a cavity 
in which the modes are sufficiently separated in 
frequency for the two-photon gain to prevail over the 
single-photon gain in an adjacent frequency corresponding 
to a dipole transition from the upper state to an 
intermediate one of opposite parity. If a cavity mode
 happens to be sufficiently near resonance with one-photon 
transition between the upper (pumped) and the intermediate 
level, the gain into that transition will take over.
The basic simple model of interaction between matter and radiation comprises two-level atoms interacting with a single-mode electromagnetic field in a lossless cavity. In spite of its simplicity, the model is intrinsically nonlinear with the atom-field coupling being the coefficient of this nonlinearity.

A brief history of the topic may be useful here. Two-photon lasers are a recurrent
 theme in the  literature and have attracted considerable theoretical 
interest semiclassically by Wang and Haken [1] as well as quantum mechanically
 [2-4]. On the experimental side, a two-photon laser has been realized and
 studied extensively in the microwave region of the spectrum [5] and in 
the optical regime by Mossberg and collaborators [6-7] in a cleaver, 
slightly different scheme, where the atomic pump transition is between 
levels of dressed atoms.
Recent work [8] in view of continuing technological improvements
in micro-cavities even at optical frequencies has motivated the
examination of certain aspects of the two-photon laser theory that are
fundamental to the process.
These aspects have their counterpart in the usual single-photon laser,
but rather different behavior is to be expected in the two-photon
case, owing to the essential nonlinearity of the process even at weak
signals.
We have here in mind a degenerate two-photon laser with the atom
pumped to the upper state connected to the lower one of the lasing
transition by a two-photon process.
Although not realized as yet in this
 pure form, it probably is a 
matter  of short time before that occurs [2-4].
 The situation here
 is somewhat different from the dressed states scheme that
 has already been demonstrated experimentally some time
 ago [6-7].
Most laser-related systems derive their feedback from a resonator structure.
  The emergence of unstable behavior in this case can be ascribed to the 
development of amplitude oscillations in an excited but previously 
stable mode or to the growth of new cavity resonances. For this reason,
 all previous investigations of this problem have focused separately 
on what can be conveniently identified as single-mode and multimode 
instabilities. Instabilities of the first type involve only the 
cavity mode that lies nearest to, but is not necessarily in 
resonance with, the atomic transition, or the carrier frequency of
 an incident field, for externally driven system. These phenomena 
can be analyzed in the context of a single-mode model. The second 
type involves the evolution of out-of-resonance cavity modes and 
therefore requires a multimode treatment for a correct description.

Most linear-stability studies avoid dealing with the spatial dependence ($\partial E/\partial z $ term in the 
field equation) [9] and
employ what has become known as the mean-field limit. By taking this limit, the field
equation is spilt into a set of purely temporal equations,
 each of which governs the evolution of a single longitudinal mode. 
Although limited in its scope, the mean-field approximation has 
proved to be quite successful in studding the stability of lasers. 
However, one should be aware that its predictions for any real 
laser need to be checked against numerical solutions of the exact Maxwell-Bloch equations.

The issue we have in mind has to do with the steady-state behavior and
its stability analysis of
the system, taking into account the  spatial 
variation 
of the cavity field along the cavity axis.
This
is most conveniently accomplished in a semiclassical formalism in 
terms of the Maxwell-Bloch
equations taking into account the spatial dependence.
Related treatments based 
on either simple rate equations [10], discussing threshold conditions,
or the Maxwell-Bloch equations without the spatial dependence, have
been presented in the literature [11-13]. What we study and present
below is essentially the generalization of the complete 
Maxwell-Bloch
equations, usually employed in the single-photon laser theory, to the
two-photon case. We have found it most convenient to use a
formulation presented some time ago by Narducci in the semiclassical
theory of the single-photon laser [14]. We present in this paper an nanlysis
which has inspired by the comparison between the linear stability analysis technique and the so-called weak sideband approach, first introduced by Casperson [15] and further elaborated by Hendow and Sargent [16] and by Boyd, Hillman, and Stroud [17]. 

\section{Derivation of equations}

Maxwell-Bloch
equations for a degenerate two-photon laser with the atom
pumped to the upper state connected to the lower one of the lasing
transition by a two-photon process and its steady state have been 
derived and discussed previously in [18].
We present 
a brief sketch of the derivation of the equations in order to specify 
the assumptions used in the derivation and in order to have at hand for
 discussion the terms relevant to the stability analysis 
of the steady state.

We consider the coupled set of Maxwell-Bloch
equations, in the usual rotating wave approximation, which govern our
two-level atom when the dipole forbidden transition is replaced by a
two-photon one. We consider the degenerate case, in which pairs of
photons with the same frequency are created or absorbed, and we
analyze the stability of the steady state.
We assume a collection of identical
homogeneously broadened 
two-level atoms, with
energies $ E_1$ and $ E_2 $ such that  ($E_2>E_1)$
with  $E_2-E_1=\hbar\omega_a $, $\omega_a $ the atomic transition frequency
 and a generated unidirectional single-mode
classical electric field
\begin{equation}
E(z,t)=\frac12 \{ E_\circ e^{i(k_cz-\omega_ct)}+c.c \}
\end{equation}
inside a ring cavity. Here $ E_\circ $ is the real field amplitude,
$k_c $  the wave-number, z  the cavity axial direction and $
\omega_c $ represents the unloaded cavity frequency.
The atoms interact with the field in the
dipole approximation via a  two-photon
transition, where  
 these states are assumed to have the same
parity, and thus are not  connected by a  one-photon
transition.

Adopting the plane-wave
approximation, we reduce the Maxwell-Bloch
equations to [18]
\begin{equation}
\frac{\partial F}{\partial z}+\frac {1}{c}
\frac{\partial F}{\partial t}=-\alpha P F^*,
\end{equation}
\begin{equation}
\frac{\partial P}{\partial t}=-(\gamma_1+i\delta_{ac})\bar
P-\gamma_1F^2D,
\end{equation}
\begin{equation}
\frac{\partial D}{\partial t}=\gamma_2\{\frac12( PF^{*2}+\bar
P^*F^2)-D+1\},
\end{equation}
where $F , P $ and $ D $ are the normalized output
field, two-photon polarization and population difference,
respectively, ($F=\sqrt{\mu^{(2)}/\hbar\gamma_1\gamma_2}\bar
E_\circ$ ),
$\mu^{(2)}$ the effective 
dipole matrix element for the two-photon transition,
$ \gamma_1 $ and $\gamma_2 $ are the decay rates of 
two-photon polarization and population difference, respectively.
$\alpha $ 
denotes the unsaturated gain constant per unit length of the 
active medium
($\alpha=2N\omega_c(\mu^{(2)})^{2/3}/2c\hbar\varepsilon_\circ
\gamma_1$ ), where N is the number of atoms per unit volume,
$ \varepsilon_\circ $ the vacuum electric permeability
and c the speed of light. We denote  by
$\delta_{ac}=\omega_a-2\omega_c $
 the detuning of the cavity mode
 from two-photon resonance.
 Maxwell-Bloch equations (2-4) have been derived by assuming an effective Hamiltonian,
i.e., by assuming a pure two-photon interaction between the two-level
atom and electromagnetic field. This approach neglects residual
effects of any largely detuned one-photon transitions between the
lasing levels and other atomic levels [19].
A more precise approach consists in assuming an exact or microscopic
interaction Hamiltonian that describes the interaction of the
electromagnetic field with a three-level cascade atomic scheme [20,21].
When the intermediate atomic level is far from one-photon resonance
the one-photon coherence can be adiabatically eliminated and the
resulting two-photon laser equations are similar to the present 
 equations
but include three additional detuning terms describing frequency
shifts. 
                     
The model is completed by appropriate boundary conditions which, in the
case of a traveling wave ring-cavity resonator, take the form 
\begin{equation}
F(0,t)=R F(L, t-(\Lambda -L)/c),
\end{equation}
where L is the length of the active medium; while the full length of
the ring resonator is $ \Lambda $ . Physically R measures the loss of the field amplitude from the exit face of the amplifying medium to its entrance face.
 
\section{Steady state}

In order to gain some physical understanding of the process and discuss some 
aspects of the threshold conditions, we analyze first the steady-state behavior of the system.
To study the steady state, we consider 
 the equations in the long-time limit by setting the time derivatives
 equal to zero,
for an  active medium  detuned by an
arbitrary amount $ \delta_{ac} $ from the resonant cavity 
mode. Under these conditions, the output field is expected to oscillate with
a carrier frequency $ \omega_L $ which is neither equal to $ \omega_c
 $ nor $ \omega_a/2 $, but to some intermediate value determined by the
 cavity and atomic parameters. 
 For this reason, we look for 
 steady-state solutions of the type
\begin{equation}
F(z,t)=F_{st}(z)
e^{-i\delta\omega t},
\end{equation}
\begin{equation}
P(z,t)=P_{st}(z)
e^{-i2\delta\omega t},
\end{equation}
\begin{equation}
D(z,t)=D_{st}(z)
\end{equation}
where $\delta\omega $ is the frequency offset of the operating laser
line from the resonant mode (i.e $\delta \omega=\omega_L-\omega_c )$.
The atomic variables can be determined at once as functions of the
stationary field profile 
\begin{equation}
P_{st}(z)=-F_{st}^2(z)
\frac{1-i\Delta}{1+\Delta^2+\mid F_{st}(z)\mid^4},
\end{equation}
\begin{equation}
D_{st}(z)=\frac{1+\Delta^2}{1+\Delta^2+\mid F_{st}(z)\mid^4},
\end{equation}
 where the detuning parameter $\Delta $ is defined as
$
\Delta=(\delta_{ac}-2\delta\omega)/\gamma_1.
$
The steady state polarization and the field envelope are generally out
of phase from one another by an amount that depends on the detuning
$\delta_{ac} $ and the position of the operating laser line. 
The steady state population difference (inversion) saturates at high intensity
levels in the sense that $ D_{st} \rightarrow 0 $ 
as $\mid F_{st}\mid\rightarrow
 \infty $. To determine the value of the output field and the form of
its longitudinal profile in steady state, it is
convenient to represent the field amplitude in terms of its modulus
only, 
\begin{equation}
\frac{d |F_{st}(z)|}{dz}=\frac{\alpha |F_{st}(z)|^3}{1+\Delta^2+ |F_{st}(z)|^4}.
\end{equation}
To make the analysis in the present paper as simple as possible we shall not consider the phase equation [18]. However, the effects of the phase variation should be taken into consideration for a mor elaborate discussion.
The boundary condition, expressed in terms of the field modulus is given by
$
F_{st}(0)=RF_{st}(L).
$
 The output laser intensity can be calculated as [18],
\begin{equation}
|F_{st}(L)|^2=\frac{2\alpha L}{1-R^2}|F_{st}(L)|^2-\frac{1+\Delta^2_j}{R^2},
\end{equation}
where
$
 \Delta_j=(\delta_{ac}-2\delta\omega_j)/\gamma_1
$, $\delta\omega_j $ is the operating laser frequency.
 Equation (12) has two roots and at laser
   threshold the intensity is not vanishing. There is coexistence of
   three solutions (above threshold): the trivial and two other
   solutions with intensity different from zero. One solution grows with
   the pumping parameter up to an asymptotic value for pumping going
   to infinity. The other solution decreases towards the zero solution as
   the pumping grows to infinity. This means that
 the threshold is not a second
   order phase transition as in the case of single photon lasers.

The quantity $c\mid ln R\mid/\gamma_1\Lambda $ represents the decay
rate of the cavity field, and $2\pi c/\Lambda $ is the spacing between
adjacent cavity resonances.
After introducing the abbreviations
$
K=c\mid ln R\mid/\Lambda,  \alpha_1=2\pi c/\Lambda,
$ we obtain 
\begin{equation}
\delta\omega_j=\omega_L-\omega_c
=\frac{K\delta_{ac}+\alpha_1\gamma_1j}{\gamma_1+2K},
\end{equation}
where the sub-index {\it j} reminds us of the possible existence of multiple
solutions. This is the well known mode-pulling formula.
It shows that the laser operating frequency is a weighted
average of the atomic resonant frequency and the frequency of one of
the cavity modes. 

\section{Linear stability analysis}

The general  stability analysis of the Maxwell-Bloch equations (2-4) 
is a rather difficult problem. The main source of
complication originates from the spatial dependence of the field and
of the atomic variables. In an attempt to get around this problem,
most linear stability analysis have been carried out within the
uniform field limit. While this may not appear to be a very realistic
approach, there are good reasons, in fact, why useful information can
be extracted even from this limiting case: (i) we can reformulate the
Maxwell-Bloch problem in terms of a new set of atomic and field
variables that are not very sensitive to limited departures from the
ideal limit. For this reason it is not necessary to operate with
unrealistically low values of the gain or the mirror transmittivity; 
(ii) the mean field limit is a good indicator of instabilities and
functions as a rough diagnostic tool. This is fortunate because the
numerical solutions of the time-dependent Maxwell-Bloch equations
require considerable efforts and some guidance can produce significant
saving of time.
The resonant case, is not very complicated and can be studied exactly with 
limited effort. For this reason, in this section we limit ourselves to the exact 
analysis of the resonant laser problem, without any restrictions on the gain of the active 
medium or the reflectivity of the mirrors. Our starting point is the full set of 
Maxwell-Bloch equations (2-4) with $\delta_{ac}=0 $. 
Because the phase of the stationary field is undetermined, it
 is possible to select $F_{st}(z) $ as a real quantity. In principle, 
a random fluctuation of the cavity field could force the growth of the
 imaginary part through a process called phase instability. 
In this section we simply assume that no phase instability
 can develop, so that both the field and polarization 
variables remain real during the linearized evolution.   
The steady state of this system of equations is given in 
equations 9, 10 and 11. To study the stability of this
steady state, we set 
\[
F(z,t)=F_{st}(z)+
e^{\lambda t}
\delta f(z),
\]
\[
P(z,t)=P_{st}(z)+
e^{\lambda t}
\delta p(z),
\]
\begin{eqnarray}
D(z,t)=D_{st}(z)+
e^{\lambda t}
\delta d(z),
 \end{eqnarray}
into Eqs. 2-4, upon neglecting fluctuation terms of order higher
than one. The linearized equation of the field fluctuation takes the
form
\begin{eqnarray}
\frac{d}{dz}\delta f(z)=M(z)\delta f(z),
\end{eqnarray}
where
\begin{eqnarray}
M(z)=-\frac{\lambda}{c}
+\frac{\lambda+3\gamma_1}{\lambda+\gamma_1}\frac{\alpha F_{st}^2}{1+F_{st}^4}- \frac{\alpha F_{st}^6}{1+F_{st}^4}
\frac{2\gamma_1+\lambda}{\lambda+\gamma_1}
\biggl(
\frac{2\gamma_1\gamma_2 }{(\lambda+\gamma_1)(\lambda+\gamma_2)+\gamma_1
\gamma_2F_{st}^4}\biggr).
\end{eqnarray}
The formal solution of Eq. 15 is
\begin{eqnarray}
\delta f(z)=\delta f(0)e^{\int_{0}^{z}dz'M(z')}=\delta f(0)e^{\Psi(z)}.
\end{eqnarray}
The problem is that $F_{st} $ is not known in closed analytic
form.
We can get around this
difficulty with a change of independent variable from z to 
 $F_{st} $, if we take advantage of the fact that $dz=
 dF_{st}/$ 
$( dF_{st}/dz) $ and that $dF_{st}/dz $ is known explicitly from
 Eq. (11)  
and the field fluctuation takes the form
\begin{eqnarray}
\delta f(z,t)=e^{\lambda t}\delta f_(z)=e^{\lambda t}\delta f(0)e^{\Psi(z)}.
\end{eqnarray}
Next, imposing the boundary condition
\begin{eqnarray}
\delta f(0,t) =R\delta f(L,t-\frac{\Lambda-L}{c}), 
\end{eqnarray}
we obtain the characteristic equation
\begin{eqnarray}
\lambda_n=-i\alpha_n 
-\frac{c}{4\Lambda}\frac{\lambda_n+2\gamma_1}{\lambda_n+\gamma_1}
\biggl[\ln\biggl(\frac{(\lambda_n+\gamma_1)(\lambda_n+\gamma_2)+\gamma_1\gamma_2F_{st}^4(L)}
{(\lambda_n+\gamma_1)(\lambda_n+\gamma_2)+\gamma_1\gamma_2R^2F_{st}^4(L)}\biggr)\biggr]
\nonumber
\\
-\frac{c}{2\Lambda}\frac{(\lambda_n+3\gamma_1)}{\lambda_n+\gamma_1}
\mid \ln R\mid,
\end{eqnarray}
where $\alpha_n=2\pi nc/\Lambda $. 
The characteristic equation (20) depends on the  cavity linewidth  K (
$c/\Lambda\gamma_1=K/\mid lnR \mid) $
of the population difference, and the gain of the active
medium through the output field intensity $F_{st}^2 $. 
The origin of the term $-i\alpha_n
$ here can be traced back to the equality $exp(0)=exp(2\pi ni) $
for
$n=0,\pm 1, \pm 2 ... . $ Note that setting  $exp(0) =1 $ would be a
mistake because it would eliminate practically the entire spectrum of
eigenvalues.
At this point, we have reduced the linearized problem (15) to the
solution of an infinite number of characteristic equations, one for
each value of $\alpha_n $.

As instructive example we wish to discuss some limiting cases of the non-linear equation (20) governing the stability of the system. 

(a) perfect reflectivity is assumed i.e $R=1$ then the only solution of equation (20) is $\lambda_n=-i\alpha_n$ i.e running waves.

(b) when the decay rate of the two-photon polarization is set to be zero, the solution of equation (20) is $\lambda_n=-i\alpha_n-\frac{c}{2\Lambda}|ln R|$ which means a stable mode.

(c) when we set the population relaxation rate $\gamma_2$ to be zero, a second order equation in $\lambda_n$ results. The solution to this equation gives the following 
acceptable formula
for the real part of  $\lambda_n$
\begin{eqnarray}
Re \lambda_n=-\frac{\gamma_1}{2}-\frac{c}{4\Lambda}|ln R|+
\sqrt{\frac{A+\sqrt{A^2+B^2}}{2}},
\end{eqnarray} 
\[
A=\frac{(\gamma_1+\frac{c}{2\Lambda}|ln R|)^2}{4}-\frac{\alpha_n^2}{4}-3\gamma_1
\frac{c}{2\Lambda}|ln R|
\]
\[
B=\frac{\alpha_n(\gamma_1+\frac{c}{2\Lambda}|ln R|)}{2}-\alpha_n\gamma_1
\]
Investigation of this equation shows that no instabilities develop in this limiting case.
This can be understood simply because writing $\gamma_2=0 $ means a constant population difference and hence no exchange of population between the two levels (see figure 1). In this figure we plot  the largest real parts of the linearized eigenvalues
 as functions of $\alpha_n $ and R. 

(d) iterative state, 
we can solve equation (20) 
by substituting the value $-i\alpha_n$ for $\lambda_n $ in the right hand side of equation (20). It can be asserted that the highly excited modes $\alpha_n $ such that $n>>10 $ are stable. This can be seen from looking at the argument of the logarithm in equation (20) in this case with the assumption $\alpha_n>\gamma_2, \gamma_1\gamma_2F_{st}^4$. However when  
$\gamma_1\gamma_2F_{st}^4>(|-i\alpha_n+\gamma_1|)(|-i\alpha_n+\gamma_2|)$ the mode becomes unstable (see figure 2). For increasing $\alpha_n$ the system becomes more stable as the positive $Re \lambda_n$-range decreases.

The existence of an infinite number of eigenvalues is not surprising
in view of the space-time dependent nature of the field and atomic
variables and of the boundary conditions of the laser resonator.
 One is reminded of the ordinary vibration problems, linear string,
 two-dimensional membrane, etc., except that here we are dealing
 simultaneously with three  fluctuation variables Eq. (14), and thus on
 physical grounds, one expects three characteristic roots
 $\lambda_n^{(1)},
 \lambda_n^{(2)},  \lambda_n^{(3)}$ for each value of n. Because $\alpha_n $ represents
 the frequency separation between the $n^{\it\underline {th}}$ empty cavity resonance and 
the selected reference mode, it is easy to interpret the set of roots 
$\lambda_n^{(i)}, i=1, 2, 3 $, as descriptive of the growth or decay of an
initial fluctuation that develops in correspondence to the $n^{\it\underline {th}}$ 
mode of the cavity. This interpretation forms the basis for a
classification of the possible unstable behaviors of the system. If
{\it Re }
$\lambda_0 $ is positive and {\it Re } $\lambda_n (n\neq 0)$ are all
negative, an initial fluctuation of the resonant mode will grow
exponentially and evolve with the same carrier frequency as the
stationary state. Thus, the linearized dynamics of the laser can be
described only in terms of the behavior of the resonant mode
fluctuation (all the other fluctuations are damped because 
 {\it Re }$ \lambda_n<0, n\neq 0$) and the instability will be  of the
 single-mode type.
If, on the other hand, {\it Re} $ \lambda_0 < 0$ and, for some value of n,
  {\it Re} $ \lambda_n >0 $, the  $n^{\it\underline {th}}$
 cavity mode will support the growth of
   a fluctuation whose carrier frequency is different from that of the
   stationary state.  Here,  the existence of a 
one-to-one correspondence between the index n, that appears in equation  
(20), and the longitudinal cavity modes is suggested. This informal suggestion is founded 
on physical grounds. The main conceptual difficulty with this interpretation
 is that the notion of "mode" is not well defined when the resonator mirrors 
have a finite reflectivity, and the elementary cavity excitations have
 a finite lifetime. In fact , in solving the linearized problem, we 
have not even introduced resonator eigenfunctions, as one normally 
would in a standered boundary value problem. For this reason, we 
continue to refer to  $ \lambda_n^{(i)}$ as the set of linearized 
eigenvalues of  the  $n^{\it\underline {th}}$ cavity resonator.

A complete analysis of equation (20), particularly with regard to the 
role played by the basic laser parameters, gain, internode spacing, reflectivity 
and the atomic decay rates, has not been carried out. Equation (20) predicts that 
both single and multimode unstable behavior can 
be established with confidence. We begin our analysis by scaling all the relevant rates of the problem to the linewidth $\gamma_1 $ of the active medium. In this way, equation (20) takes the form
\begin{eqnarray}
\bar\lambda_n=-i\bar\alpha_n 
-\frac{c}{2\gamma_1\Lambda}\frac{(\bar\lambda _n+3\bar\gamma)\mid \ln R\mid}{\bar\lambda _n+1}
-\frac{c}{4\gamma_1\Lambda}\frac{\bar\lambda _n+2}{\bar\lambda _n+1}
\nonumber
\\
\times
\biggl[\ln\biggl(\frac{(\bar\lambda _n+1)(\bar\lambda _n+\bar\gamma)+\bar\gamma F_{st}^4(L)}
{(\bar\lambda _n+1)(\bar\lambda _n+\bar\gamma)+\bar\gamma R^2F_{st}^4(L)}\biggr)\biggr]
\end{eqnarray}
where $\bar\lambda_n=\lambda_n/\gamma_1 $, $\bar\alpha_n=\alpha_n/\gamma_1 $,
and $\bar\gamma=\gamma_2/\gamma_1 $.
The equation (22) can be evaluated by using the 
iteration method as obove (in d) by substituting the value 
$-i\alpha_n $ for $\lambda_n$ in the right hand side of equation (22). It can be asserted that the highly excited modes $\alpha_n $ 
namely  $n>>10$, are stable.  
A numerical study of this problem shows that single-mode instabilities $\alpha_n=0 $
tend to be favored in the presence of high gain and laser cavity losses $K>1 $. These conditions are difficult to realize in a practical system.
In general, it appears from equation (22) that single-mode instabilities require a scaled 
cavity linewidth which is sufficiently larger than unity. 
 In order to keep the 
calculations presented in this paper as realistic as possible, we have
chosen to apply our model for a real atomic system,
(for the transition 4$p_{3/2}-6p_{3/2} $ in Potassium).
The reason for choosing this transition is the result of a
compromise. On one hand, one wants the energy of the photons involved
to be as large as possible, and preferably in the optical regime.
On the other hand, it is hard to find a two-photon transition in the
optical regime with a large coupling, since a large two-photon
coupling demands the existence of an almost resonant intermediate
level with opposite parity.
The transition mentioned above involves photons with an energy of 
$\simeq 7980 $  cm$^{-1}$ i.e. near-infrared, and has a two-photon
coupling that is orders of magnitude larger than the other candidates
we looked at, due to the almost resonant 5s state.
Besides the atom, we should also choose a cavity. In the model
presented in this paper, we are assuming that only one mode of the
cavity field is excited. For this to be true, the cavity should be
rather small, since it then supports fewer modes, and these will be
better separated in energy.
Another advantage of having a small cavity is that the two-photon
coupling $\mu^{(2)}$ will be larger, since it is proportional to
$V^{-1} $
 (following the notation of Loudon) [22], V being the cavity
 volume. We have chosen the cavity volume 
$V=10^{-15} m^3$.
Further F and $\alpha$ for the two-photon case will be proportional to $V^{-1/2}$ 
and $V^{-2/3}$ instead of $V^{-1/4}$ and $V^{-1}$ for the one-photon case. The only one that is decreasing with $(V<<1)$ very small cavities is the gain factor.

It is interesting to compare our results with those of previous work where the mean field limit is appeared. When we neglect the spatial dependence, we arrive to equation (4) of Ref. [19] but taking into account a normalization of the variables $D\rightarrow D/D_o$ , $\alpha\rightarrow kD_o$ and $P\rightarrow q/D_o$. Our formulation is based on the conventional Maxwell-Bloch equations, but is distinguished from other treatments by the inclusion of propagation effects, a finite mirror reflectivity and an arbitrary value of the gain parameter.
These features make the model more general than the previous studies.  
An example of the behavior of the eigenvalues in the case of a single-valued state equation is shown in figure 2. In this figure we plot  the largest real parts of the linearized eigenvalues
 as functions of $\alpha_n $ viewed as a continuous 
variable. For all the curves displayed in this figure we have selected 
$R=0.8 $, $\alpha_1=100 $, $\gamma=0.1 $ and for different values of $\alpha L $. We show that 
unstable situation for several values of the relevant parameters 
(the only physical meaningful values of $\bar\alpha_n $ are all the 
positive and negative multiples of the intermode spacing $\alpha_1 $).
 Multi-mode instabilities are not bounded by the high-loss requirement, but 
they still require large values of the gain to reach their threshold. 
For increasing $\alpha_n$ the system becomes more stable as the positive $Re \lambda_n$-range decreases, and the system is completely stable for $\alpha_n>10$.
In figure 3 
the largest real parts of the linearized eigenvalues are plotted 
as functions of $\alpha_n $ viewed as a continuous 
variable. With the same parameters as in figure 2 but for 
different values of $R$. This figure 
gives an example of some typical real parts of the linearized 
eigenvalues for parameter values that lead to multimode instability. 
As shown in this figure, the beat frequency due to the superposition 
of the stationary solution and of the unstable sidebands is sensitive 
to the value of $R$. The important feature is the monotonic 
shift of the positive real parts of the eigenvalues towards higher and 
higher values of $\bar\alpha_n $ for increasing values of the gain.
It is important to stress that the existence of unstable off-resonant sidebands requires a good quality cavity in the sense that $K $ must be sufficiently smaller than Unity. This is a consequence of the fact that typical instability ranges extend to maximum values of $\bar\alpha_n$ of the order of a few units. Thus, if we require that the sideband at $\bar\alpha_n=2\pi nc/L\gamma_1$ be unstable, it is necessary that the ratio $c/L\gamma_1 $ be smaller than unity. This can be arranged by selecting large enough values of $L $ or $\gamma_1$.

\section{conclusion}

We have derived the general Maxwell-Bloch equation for the system 
consisting of the two-level atoms with  dipole forbidden transition, 
placed in a two-photon one.
The treatment has been carried out in the framework of the semiclassical laser theory. 
We have calculated the spatial behavior of the field
strength 
and have shown the effect of the additional non-linearity due to the
two-photon coupling.
We have computed the 
stability analysis of the steady-state solution
of the complete 
Maxwell-Bloch. 
 Although the model is
rather idealized, its general features should be relevant to a real
single-mode system. 
The analysis presented in this paper has been inspired by the comparison between the 
linear stability analysis technique and the so-called weak sideband approach [16].
In our case the linear stability analysis not only agrees with the results of the weak sideband
approach, but extends its range of applicability, particularly in the case when the cavity 
detuning must be taken into account.

The problem we have formulated and solved in this paper has an interesting counterpart in the microwave regime where one can tailor at will, in combination with the choice of the principal quantum number of the pumped Rydberg state. The experimental realization of such a scenario should be relatively easy with present day technology. 
In our treatment we have focused on the degenerate two-photon laser.
It would thus be interesting 
to study the non-degenerate
two-photon laser. We could imagine having a transition in which one
photon is visible, and the other is, say, infrared. The frequencies of
these two photons could be chosen in such a way that we would obtain a
large two-photon coupling and hence this laser type would be easier to
realize. In this laser type would also expect Stark shift to play a
dominant role. We hope to report on such issues in a forthcoming paper.



 {\bf Captions to Figures }

~

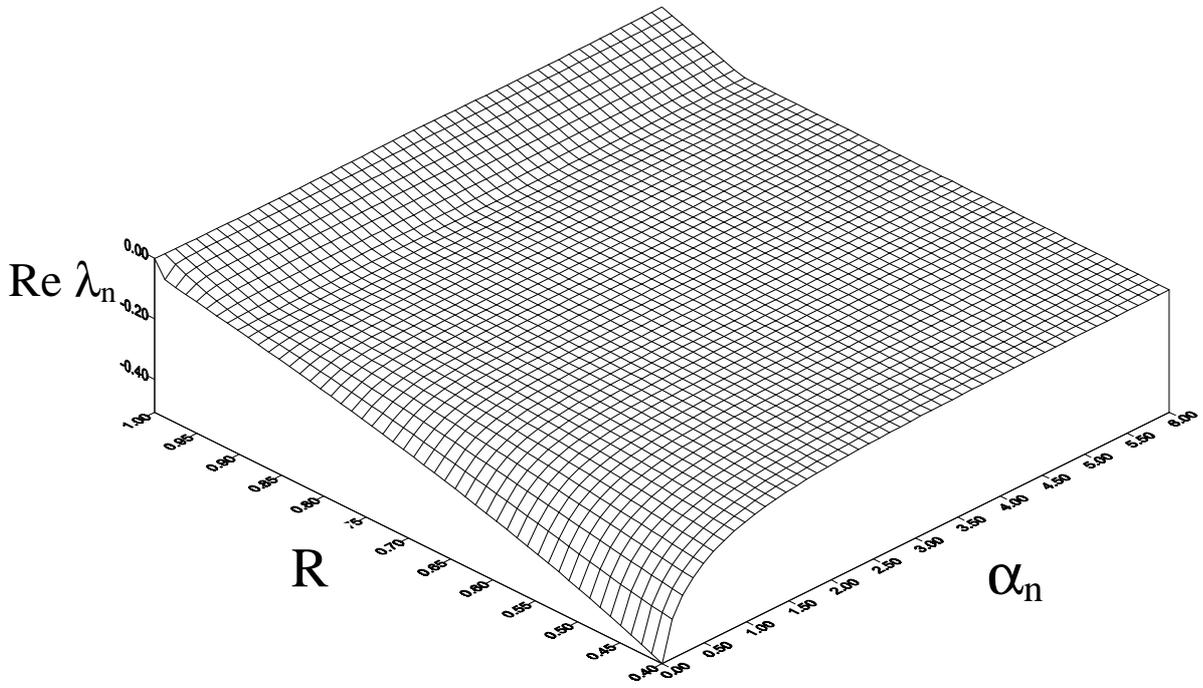
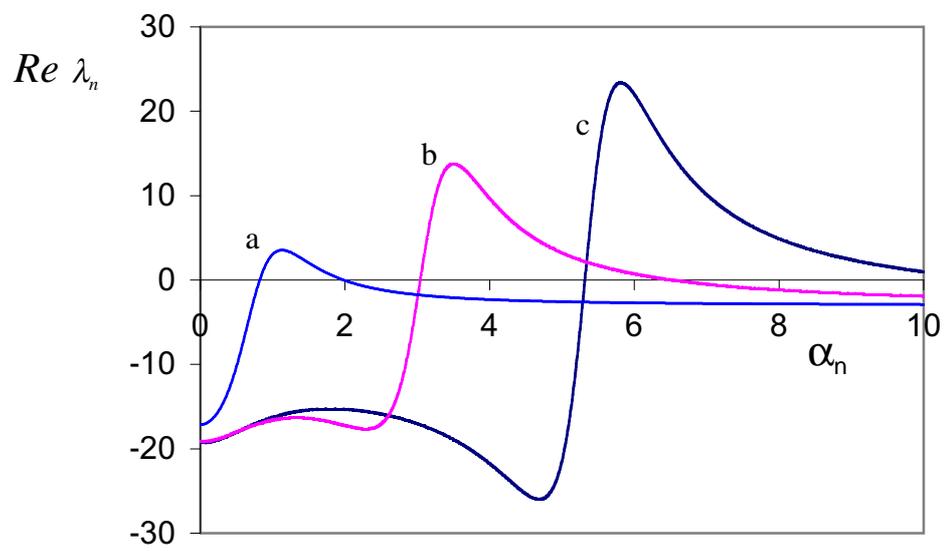
 Fig. 1: The largest real parts of the linearized eigenvalues are plotted as functions of $\alpha_n $ and R. 
We have selected $\alpha L=1$, $k=3.55 $ and for  $\gamma_1=10^5$.

~

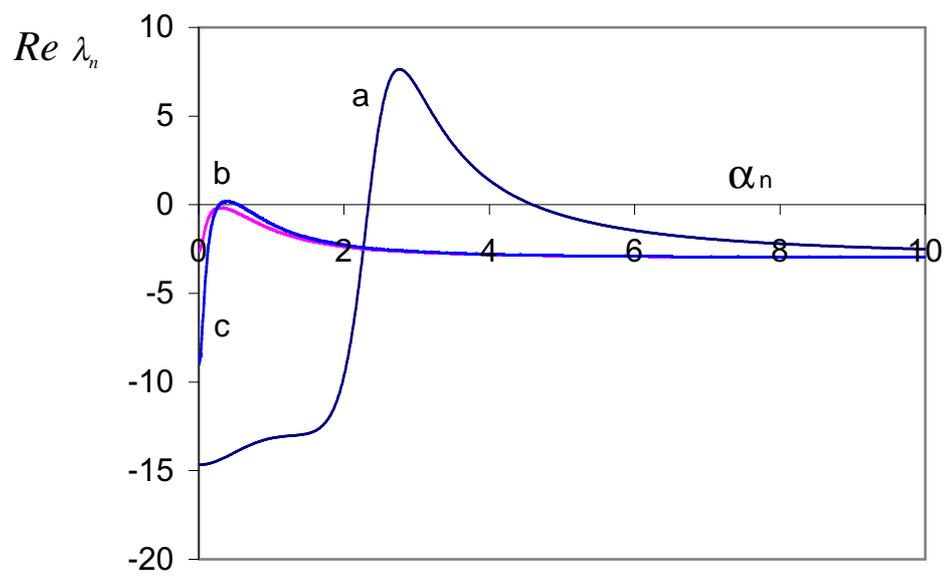
 Fig. 2: The largest real parts of the linearized eigenvalues are plotted as functions of $\alpha_n $ viewed as a continuous variable. For all the curves displayed in the figure we have selected $R=0.8, \bar\gamma=0.1, $  $k=3.55 $, and $\alpha L=1 $ (curve a), 
 $\alpha L=3 $ (curve b) and $\alpha L=5 $ (curve c).

~

 Fig. 3: The largest real parts of the linearized eigenvalues are plotted as functions of $\alpha_n $ viewed as a continuous variable. For all the curves displayed in the figure we have selected $\alpha L=5,$  $k=0.07 $ $\gamma=0.1 $,
and $R=0.3 $
 (curve a), 
$R=0.6 $
(curve b)  and $R=0.95 $
(curve c).

\end{document}